# Chirp Spread Spectrum Signaling for Future Air-Ground Communications

Nozhan Hosseini, David W. Matolak
Department of Electrical Engineering
University of South Carolina
Columbia, SC, USA 29208

*Abstract*—In this paper, we investigate the use of chirp spread spectrum signaling over air-ground channels. This includes evaluation of not only the traditional linear chirp, but also of a new chirp signal format we have devised for multiple access applications. This new format is more practical than prior multi-user chirp systems in the literature, because we allow for *im*perfect synchronism. Specifically we evaluate multi-user chirp signaling over air-ground channels in a quasi-synchronous condition. The air-ground channels we employ are models based upon an extensive NASA measurement campaign. We show that our new signaling scheme outperforms the classic linear chirp in these air-ground settings.

## I. Introduction

Aviation communication links must often be highly reliable. Even if a line of sight (LOS) component is present, depending on altitude, antenna patterns, and velocity, both air-ground (AG) and air-air (AA) channels can be dispersive and rapidly time-varying. Hence new and robust signaling schemes are of interest for aviation communications [1], since the number of flights for commercial, military, freight, and especially unmanned aerial vehicles (UAVs) is increasing.

In recent years, authors have studied performance of a variety of signaling schemes over multiple AG channels [2]-[5]. These AG channels are often classified by the environment in which the ground site (GS) is located, e.g., near-urban, or by the type of earth surface over which the flight is made, e.g., over-water, hilly.

As required link data rates increase, AG signal bandwidths also increase, and this makes signals more prone to dispersion, even in LOS cases, where multipath components (MPCs) may be from the ground, or other obstacles; see Fig. 1. Larger bandwidths are also characteristic—inherently—of spread spectrum signaling, which is often used for its ability to resolve MPCs for diversity, its robustness to narrowband interference, and the security it provides.

In this paper, we evaluate the performance of a particular type of spread spectrum (SS) on AG channels. The AG channels we employ are models developed from an extensive NASA measurement campaign [2]-[5]. The SS signal type we use is typically known as a chirp; the most common form is a linear frequency sweep over a symbol time, but we generalize this and explore performance using non-linear frequency profiles over time. We also investigate performance in multi-user conditions. Multi-user spread spectrum systems must contend with multiple-access interference (MAI). Optimal signal designs that completely eliminate MAI generally require perfect synchronization. This is often impractical, particularly for mobile platforms such as aircraft. Hence in this paper we consider AG chirp SS performance in channels that are subject to some small timing offset; we term this quasi-synchronous.

The remainder of this paper is structured as follows: in Section II we briefly summarize applications of chirp signaling. Section III contains a short description of the chirp waveforms, including a new chirp signal we have devised. For MAI estimation, we must account for the cross-correlations between signals of different users, and we illustrate this as well. In Section IV we describe the AG channels we use to evaluate our quasi-synchronous chirp SS signaling. This includes some idealized channels as well as the empirically based models. Section V contains these simulation results, and in Section VI we conclude.

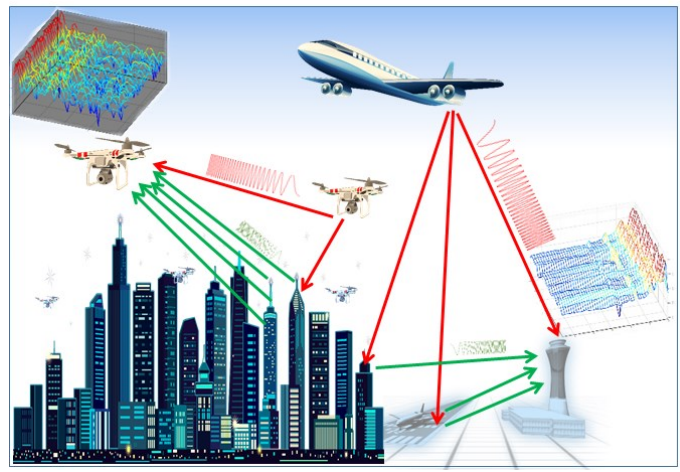

Fig. 1. Illustration of chirp signaling and air-ground channel.

## II. Chirp Signal Applications

### A. Communication

Chirp spread spectrum (CSS) signals, which can be categorized as time frequency (TF) waveforms, have several useful properties of both spread spectrum and constant envelope signals. This includes energy efficiency, and if wideband enough, robustness to multipath fading, interference, and eavesdropping. Modulation can be accomplished in several ways, with the simplest being binary sweep either up or down in frequency chirps over a bit period.. Chirps can also be used in on-off signaling or as basic waveforms for frequency shift keying (FSK). Higher order modulation can be attained with chirps in different ways e.g., by using multiple sub-bands, different start/stop frequencies (akin to pulse position modulation, PPM), and via distinct chirp waveforms within a given band.

A drawback of CSS signaling can be spectral inefficiency. This can be addressed by accommodating multiple users (waveforms) with a set of properly designed chirps in the available bandwidth. For multiple access, a set of chirp signals is required, and all waveforms in this set would ideally be orthogonal, to eliminate multiple access interference (MAI). Yet as mentioned in [6], non linear signals can also be used, and system "loading," the fraction of the total number of signals used, can also be adjusted for performance.

### B. Channel Sounding

Chirp or frequency modulated continuous wave (FMCW) can enable channel transfer function measurements. The FMCW sounder usually transmits a signal whose frequency increases or decreases linearly over a frequency range of $B$ Hz in $T$ seconds, where $B/T$ Hz/s is known as the "sweep rate" and the "time bandwidth product," or sometimes the "dispersion factor" [7].

In [8], we evaluated both "matched filter" and "heterodyne" detectors for channel sounding with chirp signals using software defined radios. The matched filter detector is based on correlating the received signal with the conjugate of the time reversed transmitted signal. Some applications also employ temporal windows to shape the correlation main lobe and side lobes, based on application.

In heterodyne detectors, the received signal is multiplied by a delayed replica of the transmitted signal and a low pass filter extracts the main lobe and other components over the desired time span. The advantage of a heterodyne detector compared to the matched filter detector is that the heterodyne detector compresses the signal in frequency instead of time, and this feature enables the use of low bandwidth digitizers and channel data acquisition equipment. Digitizer bandwidth depends on the maximum time delay or the range of the farthest multipath component.

### C. Radar

One of the earliest chirp designs was made by S. Darlington in 1947, related to waveguide transmission for pulsed radar systems with long range performance and high range resolution [9]. This technique provides a solution for the conflicting requirements of simultaneous long range and high resolution performance. These advantages provide capability of detecting a small target at greater ranges than existing systems, with better angular resolution. Similarly, many modern law enforcement agencies use FMCW radars. In [10], the authors provide a comprehensive description of using chirp for radar applications.

### III. CHIRP SIGNAL CHARACTERISTICS

As described in [6], our chirp signal's core formula is adopted from the kernel Fresnel transform theorem method. This is discussed in lightwave communication applications, and introduced in [11]. In [6] we modified the formula in a way to generate a set of $N$ orthogonal linear "up-chirps" (low to high frequency) with time (symbol) duration $T$. The $m^{th}$ waveform in complex baseband form can be written as,

$$s_m(t) = e^{\frac{j\pi}{4}} e^{\frac{j\pi N}{T^2}\left(t+\frac{mT}{N}\right)^2}, \qquad 0 \leq t < T \qquad (1)$$

where $N$ is the desired number of orthogonal chirp waveforms, $T$ is the duration of the chirp waveform and $m \in \{0, 1, \ldots N-1\}$ is the signal or user index. The total bandwidth $B$ that a set of $N$ users occupies is $B=2N/T$, and each user signal occupies the same bandwidth, $2/T$. When perfectly synchronized, the waveforms in (1) are orthogonal. A completely analogous construction can be made with "downchirps" by negating the sign of the exponent of the second term of (1). Figure 2 depicts the time-frequency (TF) plane patterns over a single symbol time. Each user signal is simply a line in the TF plane with slope $N/T$.[2]

The instantaneous frequency of the signal in (1) can be written as

$$v_m(t) = \frac{1}{2\pi} \frac{d}{dt}\left[\frac{\pi N}{T^2}\left(t^2 + \frac{2mT}{N}t + \frac{m^2 T^2}{N^2}\right)\right] = \frac{N}{T^2}t + \frac{m}{T} \qquad (2)$$

Many modern communication systems have been designed assuming quasi-synchronous conditions, where clocks of different user terminals (or, nodes) are not perfectly synchronized, but are "close" to synchronized. Their mean clock frequencies may be identical, but drift and jitter cause clocks to deviate from this mean over the short and long terms. Some detail on jitter and drift concepts is provided in [6]. In many communication systems This asynchrony is usually a small portion of a symbol duration $\delta T$ (bounded).

For the chirp waveforms of (1), the group of delayed quasi-synchronous chirp signals can be written as,

$$s_k(t) = e^{\frac{j\pi}{4}} e^{\frac{j\pi N}{T^2}\left[\left((t-\varepsilon_k)+\frac{kT}{N}\right)^2\right]}, \qquad \varepsilon \leq t < T+\varepsilon \qquad (3)$$

where $\varepsilon_k$ is the delay associated with clock drift or uncompensated propagation delay for user $k$. These delays are essentially modeled as random with some distribution, and with value limited between 0 to $T$ since other than packet transmission boundaries, effects of asynchronism recur identically over subsequent symbols. Note that complete overlap can occure for certain values of timing offset $\varepsilon_k$ that yields very large multiple access interference.

In [6], we proposed an alternative non-linear chirp signal set which, qualitatively speaking, has more "spacing" between each signal's time/frequency trace. Our heuristic approach is to fully use the available time-frequency space for signals and compare correlation performance with the linear set. Non-linear chirp waveforms can be generated with arbitrary shapes in the time/frequency domain. The most well-known examples are exponential, quadratic, and sawtooth [12], [13]. Here we propose just one specific nonlinear chirp waveform (more details are provided in [6]). Another condition we maintain is no amplitude variation. A nonlinearity function Ψ(t) is defined as in (4). This additional phase function can modify the instantaneous frequency of the linear case to any desired nonlinear TF shape. One can find the chirp signal's time-frequency shape via the time derivative $\frac{1}{2\pi}\Psi'(t)$ to find instantaneous frequency versus time.

$$s_{m_{NL}}(t) = e^{\frac{j\pi}{4}} e^{\frac{j\pi N}{T^2}\left(\left(t+\frac{mT}{N}\right)^2 + \Psi(t)\right)}, 0 \le t < T \quad (4)$$

References [12] and [13] used different mathematical derivations for their nonlinear chirp signals, but a close look at their mathematical derivation (discounting their amplitude variation) shows nonlinearity of quadratic ($t^3$) and sinusoidal (*sin(t)* and *sinh(t)*) structures, respectively.

As described previously, the intention is to increase spacing between each signal's time/frequency trace and still cover the available space. We constructed another nonlinear signal set in [6] so called "quartic nonlinear chirp". This design yields a larger time/frequency coverage than the linear case. TF plots of the linear and quartic nonlinear waveforms are shown in Figure. 2. Note that not all *N* waveforms are shown: only several of the lowest and highest frequency signals are plotted to bound each type's area. Our nonlinear case clearly occupies a larger area in the TF plane.

For the linear quasi-synchronous chirps, the cross-correlation is available in closed form [6]. Yet for arbitrary (non-linear) chirp waveforms one can generally not find any closed form solution, hence we evaluate correlations numerically.

We computed cross correlations for two signals: linear and quartic nonlinear, both in fully loaded mode. Fully loaded mode means all *N* users are sending a symbol during each time slot of duration *T*. The non-fully-loaded case has only *K* signals present, with *0<K<N*. Naturally aggregate correlations will be smaller if fewer signals are present, so the fully loaded case is the worst case. We also assume perfect power control, i.e., all user signals are received with the same power. Figure. 3 shows *average* correlation values for these two chirp types. Insets in the figure show these correlations at two smaller delay ranges, 0.05T and 0.01T, for study as QS signal candidates. We observe that the quartic nonlinear signals yield a smaller average correlation value for the entire range of timing offset for delays larger than approximately 0.02T. We also note that correlation plots are symmetric around 0.5T, therefore only delays up to this value are plotted. Additional correlation statistics have been computed [6], but are not included here for brevity.

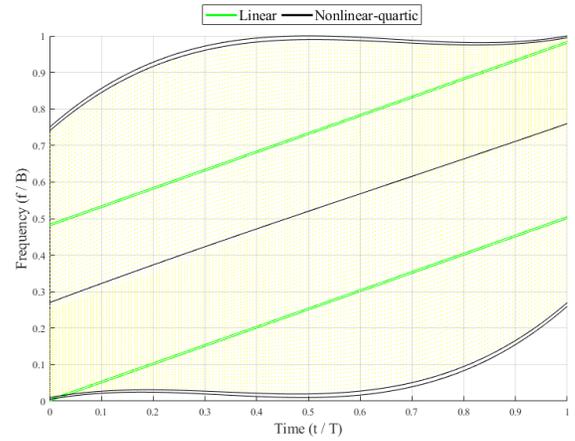

Fig. 2. Time-frequency domain representation of both linear and quartic nonlinear chirp signal sets.

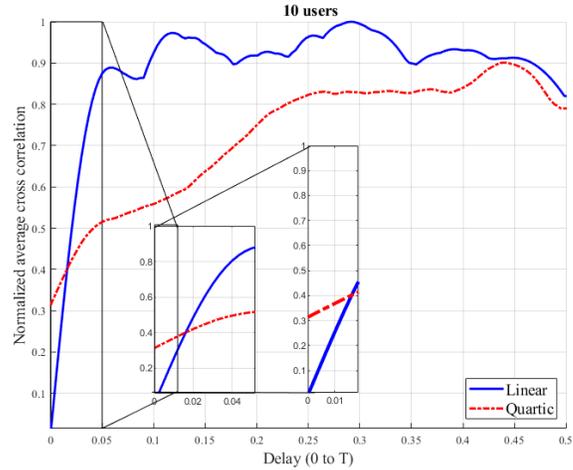

Fig. 3. Normalized average cross correlation versus delay for linear and quartic chirp set for N=10.

IV. AERONAUTICAL COMMUNICATION CHANNELS

Digital communication systems operate in specific environments, and this requires an assessment of the channel in order to ensure link reliability. One example is a high altitude air to ground aeronautical communication channel discussed in [14]. In this case the channel is considered as a free space like propagation channel, as in many satellite communication systems, with link distances larger than those in most terrestrial systems. In many terrestrial links, non-LOS (NLOS) conditions are common, and small scale fades are typically Rayleigh distributed. In contrast, in most AG channels an LOS component is present, and hence fading is Ricean distributed, with K factor in the range of 10-20 dB. This condition may not pertain throughout entire flights, of course, e.g., in future UAV links in cities where buildings can obstruct the LOS, Rayleigh fading may also occur.

Such classical flat (frequency non-selective) statistical fading models are suitable for narrowband signals, but as noted, wideband signals will encounter channel dispersion. The channel models in [2]-[5] employed a signal bandwidth of 50 MHz, in the 5 GHz spectrum. These models cover a number of different AG settings, with "medium" altitudes from 400 m to 1.7 km. The models are quasi-deterministic, based upon a detailed two-ray model, plus intermittent MPCs from ground obstacles. The 2nd "ray" is the earth surface reflection, and the intermittent MPC amplitudes, delays, and durations are modeled statistically.

## V. AG Chirp Signaling Performance

### A. Canonical Fading Channels

In this section, we investigate flat memoryless and frequency selective fading effects on the chirp spread spectrum systems that use either the classical linear chirp (1) or our proposed quartic nonlinear waveform in [6]. In the memoryless flat fading we have zero variation of the channel during a symbol time ($T_s$), but each symbol sees a different realization of the fading distribution. In general this condition is not quite realistic for most channels where coherence time $t_c$ is larger than the symbol time ($t_c >> T_s$), but can be approximated in such slowly fading channels via interleaving. In addition, analytical performance evaluation is also often possible in this case. In contrast, in our fast fading channel there *is* variation within a symbol time. Both effects are illustrated for CSS waveforms in Fig. 4.

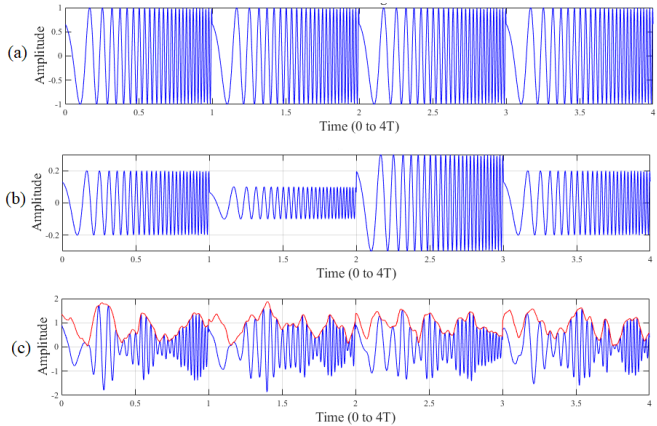

Fig. 4. Time domain representation of CSS signals in (a) no fading, (b) memoryless flat fading, and (c) fast (freq-selective) fading, for $f_D T$=0.01.

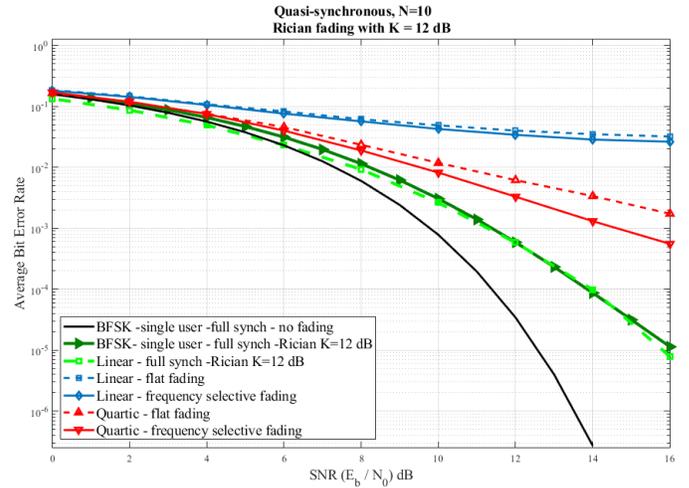

Fig. 5. BER vs. SNR chirp performance in Ricean fading (K=12 dB) performance, in quasi-synchronous conditions ($\sigma = 0.1T$).

Fig. 5 shows bit error ratio (BER) versus received signal to noise ratio (SNR) in terms of the energy per bit to noise density ratio $E_b/N_0$ for several cases. Signaling for the chirps is binary, with slope (up or down) denoting the two bits, and the number of users is N=10 for all cases except the black line that represents the single-user, no fading (additive white Gaussian noise) channel. Signal bandwidths of the systems are equal (20/T). No equalization is used, and for the chirps we distribute delays randomly with a zero-mean Gaussian distribution, standard deviation 0.1T. As we can see in Fig. 5, the fast fading with $f_D T$=0.01 has slightly better BER performance than our memoryless Ricean case, for quartic case more than linear case. The fully synchronous linear case (achieving orthogonality) is also shown for comparison. Most notably, the quartic chirp waveforms outperform the linear case [6].

### B. Empirical AG Channel

As noted, references [2]-[5] provide several empirical AG channel models for use in evaluation of UAV communication systems. Reference [4] specifically addresses the hilly suburban setting, which we use here. This AG channel is represented as a two ray model with Ricean fading, with mean K-factor of 12 dB (for L-band), along with 3rd to 6th rays that are statistically modeled intermittent MPCs. Based on the results in [4], two cases of mean and worst-case delay spreads can be modeled. In Figures 6 and 7 we depict example hilly suburban AG channel power delay profiles versus time for the mean and worst-case, respectively. We can see the delay spreads in Fig. 7 are as expected larger than those in Fig. 6.

Fig. 8 plots the BER vs. SNR performance results for these channels for our designated system in [6] with $T$=10 μs, corresponding to $B$= 2 MHz for 10 users system, and with the zero-mean Gaussian random timing offsets of two different standard deviation values. Once again, the quartic chirps outperform the linear chirps in this realistic AG channel, with practical timing offset conditions.

## VI. CONCLUSION

In this paper, we briefly discussed chirp signaling for use in aeronautical communication. We described the traditional linear chirp, and introduced a new chirp design that is advantageous in practical non-perfect "quasi-ynchronous" conditions. We simulated performance in some "canonical" Ricean fading channels, and in realistic aeronautical channels based on extensive measurements described in [4]. The BER performance of our proposed quartic chirps is superior to that of the linear chirps for these practical AG channels. Future work will investigate performance in additional AG channels and will investigate additional non-linear chirp designs.

## I. ACKNOWLEDGEMENT

The authors would like to thank Dr. H. Jamal of the University of South Carolina for development of the AG channel Matlab routines.

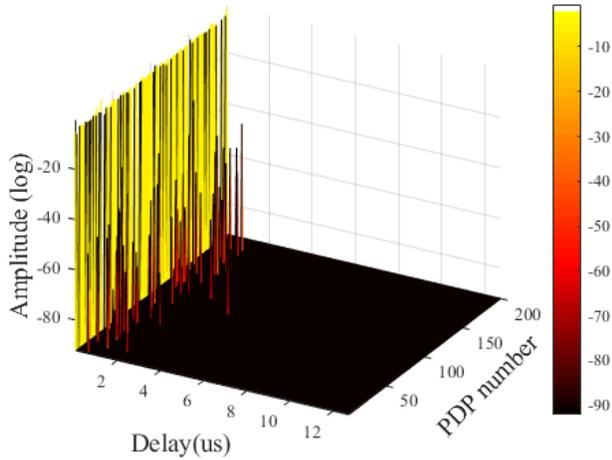

Fig. 6. Mean channel PDPs vs. time for empirical AG hilly suburban model in [4].

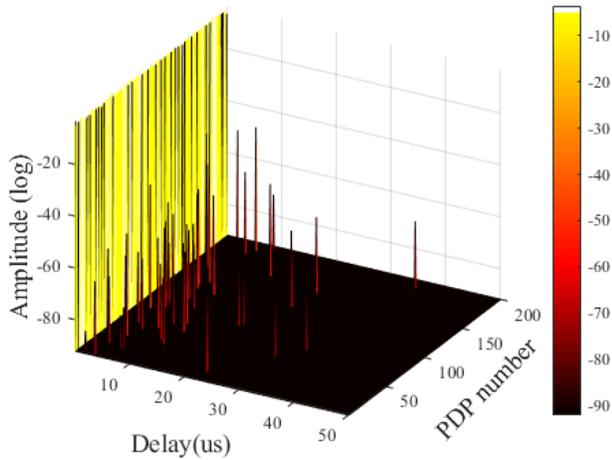

Fig. 7. Worst case channel PDPs vs. time for empirical AG hilly suburban model in [4].

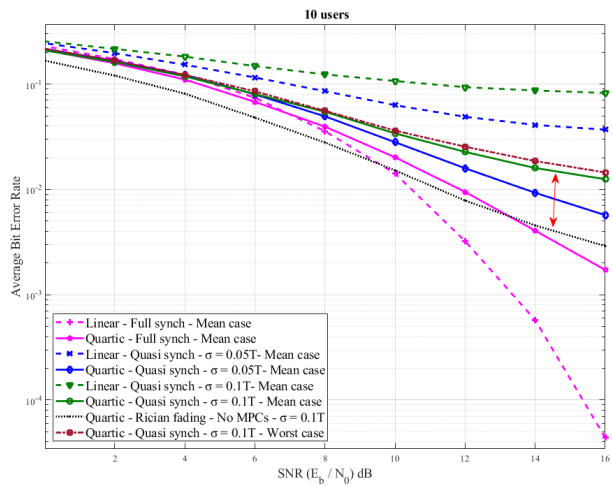

Fig. 8. BER vs. SNR for chirp signaling in quasi-synchronous conditions, over AG hilly channel model of [4].